\documentstyle[12pt]{article}

\def\Fig{{\bf Fig.}}

\def\P{{\bf \Pi}}

\begin{document}
\begin{flushleft}
\vspace{-1 cm}
NEIP-96-005
\end{flushleft}
\begin{flushright}
\vspace{-1cm}
\today \\
\vspace{2 cm}
\end{flushright}
\begin{center}
\Large
{\bf  On the Convergence to Ergodic Behaviour of Quantum Wave Functions} \\
\vspace{0.5 in}
\large
Ph. Jacquod \footnote{e-mail philippe.jacquod@iph.unine.ch}
 and J.-P. Amiet \footnote{e-mail amiet@iph.unine.ch}\\
\vspace{0.1in} Institut de Physique \\
Universit\'e de Neuch\^atel \\
1, Rue A.L. Breguet  \\
CH - 2000 Neuch\^atel \\
\end{center}
\normalsize
\vspace{0.2in}
\begin{center}
{\bf Abstract }
\end{center}
We study the decrease of fluctuations of diagonal matrix elements of
observables and of Husimi densities of quantum mechanical wave functions around
their mean value upon approaching the semi-classical regime ($\hbar \rightarrow
0$). The model studied is a spin (SU(2)) one in a classically strongly chaotic
regime. We show that the fluctuations are Gaussian distributed, with a width
$\sigma^2$ decreasing as the square root of Planck's constant. This is
consistent with Random Matrix Theory (RMT) predictions, and previous studies on
these fluctuations \cite{eck,fein}. We further study the width of the
probability distribution of $\hbar$-dependent fluctuations and compare it to
the Gaussian Orthogonal Ensemble (GOE) of RMT.

\newpage
\noindent The behaviour of quantum mechanical wave functions in the
semiclassical limit has recently attracted much interest. It is motivated by
the fact that
the spectrum alone cannot contain the whole information on the system. Roughly,
one can say that in integrable systems the eigenfunctions condense on
classically invariant
torii, while in chaotic ones, where such classical structures have been
destroyed, they tend to spread uniformly over the whole classically allowed
region. Few
analytical results have been obtained however in chaotic regimes, the most
important of which perhaps is the Shnirelman theorem. One formulation of this
theorem would be that in the limit $\hbar \rightarrow 0$, almost all the
diagonal matrix elements of almost all quantum mechanical observables converge
weakly to a constant over the classically chaotic region \cite{shni}. A few
years ago Feingold and Peres \cite{fein} and more recently, Eckhardt et. al.
\cite{eck} have studied the rate of this convergence for autonomous systems
where the semiclassical limit is, according to the Shnirelman theorem, the
microcanonical phase-space (i.e. classical) average. As they mentioned, the
"almost all quantum mechanical observables" in this formulation exclude
projection operators, and in general all operators without smooth classical
limit. Moreover, the "almost all diagonal matrix elements" still leave room for
scarring of eigenstates by short periodic orbits \cite{hel}. For those states,
the limit can be dramatically different from the Shnirelman-predicted one.
Their conclusion
is that in a strongly chaotic system and for a smooth classical observable
$A(p,q)$ with which a quantum operator $\hat{A}$, $
A_{jk}:=<j|{\hat A}|k>$, can be associated,  the fluctuations of the diagonal
matrix elements
\begin{eqnarray}
<F^2_j>:=<(
A_{jj}-\{A\})^2>
\end{eqnarray}
around the semiclassical microcanonical average
\begin{eqnarray}
\{A\} = \int A(p,q) \delta(E-H(p,q)) d^dp d^dq/\int \delta(E-H(p,q)) d^dp d^dq
\end{eqnarray} have the
same order of magnitude as the mean square of the off-diagonal terms $<|
A_{jk}|^2>$ and decrease proportionally to the inverse of the Heisenberg time
$1/T_{H} \sim \hbar$ upon approaching the semiclassical limit, in agreement
with Random Matrix Theory (RMT) predictions.
They related the proportionality
coefficient to the autocorrelation function of the classical dynamical variable
$A$, $C(t):= \lim_{T \rightarrow \infty} {\frac{1}{T}}
\int_{0}^{T} A(t+\tau)A(\tau) d\tau $,
i.e.
\begin{eqnarray}
<F^2_j> = \frac{2}{T_H}\int_{0}^{\infty} dt C(t)
\end{eqnarray}
In particular, almost all diagonal elements $A_{jj}$ tend to the semiclassical
microcanonical average as $\hbar \rightarrow 0$.
Eq.(3) states among others that {\it quantum fluctuations are proportional to
classical correlations}. Their argument goes as follows : According to
Shnirelman's theorem, the diagonal matrix element
\begin{eqnarray}
<E_j|\hat{A}(t) \hat{A}(0)|E_j> \rightarrow C(t) \hspace{2cm} \hbar
\rightarrow 0
\end{eqnarray} On the other hand, this matrix element is
\begin{eqnarray}
<E_j|\hat{A}(t) \hat{A}(0)|E_j> & = &\sum_k \exp
\left[i(E_j-E_k)t/\hbar\right] |A_{jk}|^2 \nonumber \\
& = & \sum_{k \neq j} \exp
\left[i(E_j-E_k)t/\hbar\right] |A_{jk}|^2 + |A_{jj}|^2
\end{eqnarray}
 Thus we have
\begin{eqnarray}
\sum_{k \neq j} \exp \left[i(E_j-E_k)t/\hbar \right] |A_{jk}|^2 \rightarrow
C(t) - \{A\}^2 \hspace{2cm}\hbar
\rightarrow 0
\end{eqnarray}
Defining the Fourier Transform of the autocorrelation function $S(\omega) :=
\int_{-\infty}^{\infty} C(t) \exp(-i \omega t) dt $ we
have
\begin{eqnarray}
|A_{jk}|^2 \approx S((E_j-E_k)/\hbar)/(2 \pi \rho(E)) = \int_{-\infty}^{\infty}
\left[C(t)-\{A\}^2\right]dt  \hspace{1cm} E_j \rightarrow E_k
\end{eqnarray}
Then, under the assumption that as $E_j \rightarrow E_k$, the eigenfunctions
$|E_j>$, $|E_k>$ and $|\pm>:=\frac{1}{\sqrt{2}} (|E_j> \pm |E_k>)$ are
qualitatively similar, i.e. :
\begin{eqnarray}
A_{jk} \approx <-|\hat{A}|+> \nonumber \\
\{A\} \approx <+|\hat{A}|+> \approx <-|\hat{A}|->
\end{eqnarray}
we have
\begin{eqnarray}
A_{jk} \approx <-|\hat{A}|+> = \frac{1}{2} (A_{jj}-A_{kk}+A_{jk}-A_{kj})
\end{eqnarray}
Finally defining the fluctuations as $F_j :=A_{jj}-\{A\}$ and assuming
statistical independence of the $F_j$'s, i.e.: $<F_j^2> = <F_k^2> = 2
<|A_{jk}|^2>$ we get eq. (3).
Illustrations of this result on the double rotator model \cite{fein}, the
bakers map and the hydrogen atom in a magnetic field \cite{eck} nicely
confirmed
these predictions. These are to our knowledge the only works that delt with the
qualitative description of the approach to ergodicity of quantum mechanical
wave functions. Here, we extend these results to a kicked (e.g. non autonomous)
system. We will focus on the fluctuations of the Husimi density of the
eigenstates, i.e. study the fluctuations of the diagonal matrix elements of the
projection operator over coherent states \cite{perelo}. The Hamiltonian
\begin{eqnarray}
H := \frac{\hbar}{4 S T} S_{z}^{2} + \frac{\hbar \kappa}{T}
S_{y} \sum_{n=-\infty}^{+\infty} \delta (t-nT)
\end{eqnarray}
is expressed in term of the SU(2) spin operators $S_x$, $S_y$ and $S_z$,
while $0 \leq \kappa\leq 2 \pi $. Models of this kind have been extensively
studied \cite{jac} and are usually referred to as "kicked tops". They
alternatively represent a
spin which evolves during a time $T$ under the influence of an integrable
hamiltonian after which it undergoes a rotation of angle $\kappa$ around the
y-axis. It thus defines the time evolution (Floquet) operator :
\begin{eqnarray}
U_T:= \exp(-i \frac{\kappa}{T} S_{y}) \exp(-\frac{i}{4 S}
S_{z}^{2} )
\end{eqnarray}
The above argument leading to eq. (3) must be slightly modified in order to
apply to the map defined by eq.(10) and eq.(11). Instead of working with energy
eigenstates $|E_j>$ of an autonomous Hamiltonian, we deal with quasienergy
eigenstates $|\omega_j>$ of an unitary time evolution operator. As a
consequence, the microcanonical average of eq.(2) is replaced by a phase space
integral restricted to the corresponding connected chaotic region. In our case
and in a strongly chaotic regime eq.(2) reads :
\begin{eqnarray}
\{A\} = \int_{{\cal S}^2}A(\theta,\phi) \sin(\theta)d\theta d\phi / \int_{{\cal
S}^2} \sin(\theta) d\theta d\phi = \frac{1}{4 \pi} \int_{{\cal S}^2}
A(\theta,\phi)  \sin(\theta) d\theta d\phi
\end{eqnarray}
i.e. we integrate over the whole sphere ${\cal S}^2$ instead of the energy
surface. In the semiclassical limit, the diagonal matrix elements
\begin{eqnarray}
<\omega_j|\hat{A}(t) \hat{A}(0)|\omega_j> = \sum_k \exp
\left[i(\omega_j-\omega_k)t/\hbar\right] |A_{jk}|^2 \rightarrow C(t)
\end{eqnarray}
provided the regime studied is classically strongly chaotic. Moreover, a
similar argument as before leads to
\begin{eqnarray}
|A_{jk}|^2 \approx S((\omega_j-\omega_k)/\hbar)
\end{eqnarray}
and hence we recover eq.(3). Here, we concentrate on the study of the
eigenstates of the unitary operator eq.(11)  in the regime $T=50$ and $\kappa =
1.2$. By standard numerical computation of the
Lyapounov exponent \cite{lili} over the whole phase space, we checked that in
this
regime the classical motion is strongly chaotic. Moreover, we checked that the
quantum mechanical operator eq.(11) exhibits the usual characteristics of
quantum
chaos : its level spacings statistics and spectral rigidity follow the
predictions of the Circular Orthogonal Ensemble (COE) of RMT.

\noindent As mentioned in \cite{eck} the Shnirelman theorem leaves room for
wave functions to show large deviation from the semiclassical limit value. It
only states that the proportion of such wave functions should be negligible,
i.e. in
the semiclassical limit, they build a subset of zero measure.
For "almost all eigenfunctions" then, the variance of these fluctuations should
vanish as $\hbar \rightarrow 0$. However, this decay can be substantially
perturbed by scarring of eigenfunctions by short periodic orbit \cite{hel} :
Scarred eigenfunctions are front-line candidates for exceptions to the
Shnirelman theorem ! Thus they could significantly -  and negatively - affect
our results.
It has been suggested that scarring manifests itself in deviations of RMT
predictions in the level curvature distribution \cite{del} \cite{tah}. Though
not yet rigorously proven, this statement is now widely accepted. This
distribution for the model defined by eq.(11) in the regime studied is shown in
fig.1.
The remarkable agreement with RMT (full curve) prediction indicates a small
number of scarred eigenstates, an agreement which was already obtained on a
similar model \cite{del}. Therefore, scarring is not likely to influence our
study.

\noindent Let us briefly outline our method. Our goal is to study the behaviour
of eigenstates of eq. (11) in the semiclassical limit, i.e. as $\hbar
\rightarrow 0 $, $S
\rightarrow \infty$ so as to leave the product $\hbar S$ constant.
A peculiarity of such systems is that the parameter governing the convergence
to the semiclassical limit governs too the number of states $2S+1 \sim 1/\hbar$
and the density of states. In order to determine the implication of this
peculiarity on our study, we will therefore check the validity of our results
on GOE matrices.

\noindent The Husimi density of an eigenstate $|\omega>=\sum_{\mu=-S}^{S}
\omega_\mu |\mu>$ of eq.(11) is defined as the projection of this state onto a
coherent state $ |\theta,\phi> $ of the spin SU(2) group \cite{perelo} :
\begin{eqnarray}
&  \Omega_{\omega}^S (\theta,\phi):=|<\omega|\theta,\phi>|^2 \nonumber \\
& |\theta,\phi> := \sum_{\mu=-s}^{s}\sqrt{\left(^{\hspace{0.15cm}2 s}_{s-\mu}
\right)} \sin(\frac{\theta}{2})^{s-\mu} \cos(\frac{\theta}{2})^{s+\mu} e^{i
(s-\mu) \phi} |\mu >
\end{eqnarray}

\noindent The $\Omega_{\omega}^S$ are smooth functions of $\theta$ and $\phi$
and thus
can be expanded in term of the spherical harmonics :
\begin{eqnarray}
&  \Omega_{\omega}^S (\theta,\phi) = \sum_{l,m} {\sqrt{\frac{4 \pi}{2l+1}}
\Omega_{l,m}^S Y_{l,m} (\theta,\phi)}
\end{eqnarray}
where $l=0,1,2,...2S$ and $m=-l,-l+1,-l+2,...l$. We used the convention to
introduce the square root in this expansion. This multipole expansion allows us
to
interpret the $\Omega_{l,m}^S$ in term of magnitude of fluctuations of size
$\sim \frac{\pi}{m+1}$ in the $\phi$-direction and $\sim \frac{2 \pi}{l+1}$ in
the $\theta$ direction. We will thus get quantitative results on the decrease
of fluctuations as a function of their size. Let us recall that the Shnirelman
theorem implies that as $\hbar = 1/S \rightarrow 0$, fluctuations of fixed and
non zero $l$ must vanish. However it does not say anything about the behaviour
of, say, $\Omega_{l(S),m(S)}^S$ as $ S \rightarrow \infty$ when $l(S)$ and
$m(S)$ are monotonously increasing functions of $S$ , i.e. investigating such
moments could lead us to different conclusions than that of \cite{eck,fein}.

\noindent Using the resolution of unity :
\begin{equation}
{\bf 1} = \frac{2 s+1}{4 \pi} \int d\theta d\phi \sin \theta
|\theta,\phi><\theta,\phi|
\end{equation}
the normalization condition reads :
\begin{eqnarray}
& 1 &   = <\omega| {\bf 1}|\omega> = (2S+1) \Omega_{0,0}^S \nonumber \\
& & \Rightarrow  \Omega_{0,0}^S  = \frac{1}{4 (2S+1)}
\end{eqnarray}

\noindent i.e. the $0^th$ moment decrease as $1/S \sim \hbar$ on approaching
the semiclassical limit. In the following, we therefore divide all higher
moments $\Omega_{l,m}^S$ by $\Omega_{0,0}^S$ to consistantly study their
decrease and introduce the notation ${\hat
\Omega}_{l,m}^S:=\frac{\Omega_{l,m}^S}{\Omega_{0,0}^S}$.
Let us note that this $1/S$ behaviour of the Shnirelman limit is a consequence
of the overcompleteness of the coherent states representation. On the other
hand
we have :
\begin{eqnarray}
{\hat \Omega}_{l,m}^S  = 4(2S+1)\sum_{\mu=-s}^{s} \omega_{\mu}^* \omega_{\mu+m}
(-1)^{s-\mu}
C_{s,-s,0}^{s, s, l} C_{\mu+m,-\mu,m}^{s, s, l}
\end{eqnarray}
which gives a check of our numerical computation for small $S$. However, the
numerical difficulty for computing the Clebsch-Gordan coefficients $C_{\mu+m,
-\mu,m}^{s,s,l}$ for large $S$ leads us to use the following numerically more
stable
and faster method to compute the moments ${\hat \Omega}_{l,m}^{S}$. We define :
\begin{eqnarray}
M_{k,m}^S (\omega) := 4(2S+1)<\omega| S_z^k S_-^m|\omega>/S^{k+m}
\end{eqnarray}
It is straightforward to see that there is a linear relation between the
$M_{k,m}^S$ and the ${\hat \Omega}_{l,m}^S$ : (We use the shorter notation
$\gamma=e^{i \phi} \tan(\theta/2)$)
\begin{eqnarray}
& M_{k,m}^S (\omega) & = \frac{1}{S^{k+m}} Tr \left[ |\omega><\omega| S_z^k
S_-^m \right] \nonumber \\
& & = \frac{2S+1}{4 \pi S^{k+m}} \int {d^2 \gamma}   <\gamma|\omega><\omega|
S_z^k S_-^m|\gamma> \nonumber \\
& & = \frac{2S+1}{4 \pi S^{k+m}} \int {d^2 \gamma}  {\hat \Omega}_{\omega}^{S}
(\gamma) \circ ({\cal S}_z \circ)^k ({\cal S}_- \circ)^m
\end{eqnarray}
The curved letters ${\cal S}$ stand for classical quantities, and for any
function $f(\gamma)$ we have defined the product \cite{amci}:
\begin{eqnarray}
f(\gamma) \circ {\cal S}_z := ({\cal S}_z - \gamma {\partial \over \partial
\gamma}) f(\gamma) & & \nonumber \\
f(\gamma)\circ {\cal S}_- := ({\cal S}_- + {\partial \over \partial \gamma})
f(\gamma) \nonumber
\end{eqnarray}
This allows us to write $ <\gamma|\omega><\omega| S_z^k S_-^m|\gamma>$ as a
differential operator acting on ${\hat \Omega}_{\omega}^{s}(\gamma)$. The trick
is
then to partially integrate this expression. After a little algebra we reach :
\begin{eqnarray}
 & & \frac{2S+1}{4 \pi S^{k+m}} \int {d^2 \gamma}  {\hat \Omega}_{\omega}^{S}
(\gamma)
\circ ({\cal S}_z \circ)^k ({\cal S}_- \circ)^m \nonumber \\
& = & \frac{(2S+1+m)!}{(2S)! 2^{k+m} 4 \pi}  \int_{0}^{2 \pi}{d \phi }
\int_{-1}^{1}{d u}e^{-i m \phi} (1-u^2)^{m/2} {\cal P}_{k,m}^{S} (u)
{\hat \Omega}_{\omega}^{S}
\end{eqnarray}
where
\begin{eqnarray}
{\cal P}_{k,m}^{S} (u)  = ((2S+2+m) u - m
-(1-u^2){d \over du})^k 1 := \sum_{l^{'} = m}^{k+m} p_{k}^{l'} P_{l'}^{m}(u)
\end{eqnarray} in
term of the Legendre polynomials $P_{l'}^{m}(u)$ and $u = \cos(\theta)$. We
finally get :
\begin{eqnarray}
& M_{k,m}^S (\omega) & = \frac{(2S+1+m)!}{(2S)! (2S)^{k+m}} \sum_{l=m}^{k+m}
\frac{1}{2l+1} {\hat \Omega}_{l,m}^{S} p_{k}^{l}
\end{eqnarray}
It is thus possible to obtain the ${\hat \Omega}_{l,m}^{S}$ through a matrix
multiplication of the moments $ M_{k,m}^S (\omega) $ :
\begin{eqnarray}
M^S (\omega) = {\cal M} {\hat \Omega}^S
\end{eqnarray}
where we defined $(M^S (\omega))_{k,m}= M_{k,m}^S (\omega)$, $({\hat
\Omega}^S)_{l,m} = \frac{(2S+1+m)!}{(2S)^{m}}{\hat \Omega}_{l,m}^{S}$ and
$({\cal M})_{k,l} =\frac{1}{(2S)! (2S)^{k}}
\frac{1}{2l+1}  p_{k}^{l}$.

\noindent Numerical invertion of this last matrix
allows us to get the multipoles ${\hat \Omega}_{l,m}^S$ from the numerical
computation
of the moments $ M_{k,m}^S (\omega) $. The advantage of this method against the
direct computation of Husimi densities is the numerical stability. Moreover, if
we are interested in the few first moments, say up to $l \ll S$, then only the
diagonal matrix elements up to $M_{l,m}^S$ are necessary.

\noindent Fig.2 shows a plot of a moment distribution $P({\hat
\Omega}_{1,0}^{600})$
obtained through computation of 2404 diagonal matrix elements from 4 unitary
matrices defined by eq.(11) \footnote{We have considered only the projection of
(2) on even
states, i.e. states which are left invariant by the parity $ \P | \mu > = |-
\mu> $.} close to the regime $T=50$ and $\kappa=1.2$. The agreement with
the Gaussian fitting is remarkable and allows us to conclude that the
fluctuations of the ${\hat
\Omega}_{l,m}^S$ obey the probability
distribution
\begin{eqnarray}
P({\hat
\Omega}_{l,m}^S) \propto \exp(-( {\hat
\Omega}_{l,m}^S- {\hat
\Omega}^\infty_{l,m})^2/(2
\sigma_{l,m}^2))
\end{eqnarray}
where the mean value $ {\hat \Omega}_{l,m}^{\infty} $ is the Shnirelman limit.
This
distribution narrows itself as $\hbar \rightarrow 0$, until finally the "almost
all" wave functions, i.e. those who obey the Shnirelman theorem, have converged
to their semiclassical shape $\hat{\Omega}_{l,m}^\infty = 0$, $l \neq 0$ and $
\hat{\Omega}_{0,0}^\infty  = 1$. In other words, $\sigma_{l,m}^2$ decreases as
$S$
increases. This decay follows a power law as shown in Fig.3. We have :
\begin{eqnarray}
\sigma_{l,m}^2 \sim S^{-1/2} & & \forall l \neq 0
\end{eqnarray}
As already mentioned, this law is valid for fixed $l$ and $m$ in the regime
$l$, $m \ll S$.

\noindent We further did the same study on GOE matrices. The result is shown in
Fig.4 and indicates a decay of the width of the Gaussian distribution of
fluctuations of the Husimi density of the form (25). Let us note at this stage
that the relationship between this width and the fluctuations studied in
\cite{eck,fein} is :
\begin{eqnarray}
(\sigma_{l,m}^2)^2 \sim <F^2_j>
\end{eqnarray}
We thus get the same $1/S$ decay of the fluctuations. As for the shape of these
fluctuations, the
diversity of models studied up to date leads us to postulate that {\it quantum
mechanical
systems with strongly chaotic classical counterpart have gaussian distributed
fluctuations of their diagonal matrix elements around their microcanonical
classical average (Eq.(2) or eq.(12)).} Apparently, the width of this gaussian
decays like $\hbar$ as $\hbar \rightarrow 0$. This postulate is to be
taken with the "almost all" Shnirelman restrictions and excludes of course
models like the kicked rotator \cite{she}, where quantum interference effects
lead to localization of the wave function, thus destroying the ergodicity of
the quantum wave function \footnote{However restriction of quantum means to
phase space region smaller than the localization length should lead to a
similar behaviour.}.

\noindent Up to now we have shown that our model matches in every respect all
the features of a GOE random matrix : its spectrum exhibits level repulsion its
level curvature statistics correspond to the RMT predicted distribution and the
statistical distribution of the components of its eigenvectors tends to the
semiclassical average in the same way, which in its turn implies a decay of the
width  of the Gaussian distribution of the moments ${\hat \Omega}_{l,m}^S$
defined in (16). However as has already been said, there is absolutely no
reason to expect a similar decrease of such moments when $l$ is not small
compared to $S$. We therefore turn our attention to the behaviour of such
moments.

\noindent We concentrate on the questions :

$\bullet $ Is there a similar power-law decay for $\hat{\Omega}^S_{l(S),m(S)}$
when $l(S)$ and $ m(S)$ are increasing functions of $S$ ?

$\bullet $ Are there possibly restrictions on $l(S)$ and $ m(S)$ for this
power-law to stay valid ?

\noindent Answering this questions gives us information on the minimal size
$\Delta_{l,m}$ of the relevant fluctuations. From the Heisenberg uncertainty
principle we have a lower bound for the fluctuations size $\Delta_{l,m}
=\frac{2 \pi^2}{(l+1) (m+1)} >_{\hspace{-0.25cm} \sim} \hbar \sim 1/S$ and thus
an upper bound for $l$ and $m$ : $l,m {\
\lower-1.2pt\vbox{\hbox{\rlap{$<$}\lower5pt\vbox{\hbox{$\sim$}}}}\ } S$. For
the sake of simplicity we will restrict ourselves to the study of $m=0$
moments, and will study moments with $l \sim S$ and $\sqrt{S}$ using formula
(19) with random eigenfunction components $\omega_{\mu}$ which corresponds to
the GOE case \footnote{The $\omega_{\mu}$'s are random up to the normalisation
condition $\sum_{\mu=-S}^S |\omega_{\mu}|^2 = 1$ and the $\Pi$-parity :
$\omega_{\mu} \neq 0$ either for $\mu=-S$, $-S+2$, $-S+4$, ... $S$ or
$\mu=-S+1$, $-S+3$, ... $S-1$.}.

\noindent We show the result of this study on fig.5 for $l(S) = S/2$, $3S/4$,
$S$ and $5S/4$. Obviously, these $S$-dependent moments decay faster than those
with fixed $l$ and $m$. Moreover a $S_c$ is likely to exist for each $l(S)$
above which the magnitude of the corresponding fluctuation decays faster than a
power law, possibly exponentially. However this latter conclusion is to be
taken carefully because of the restricted $S$-range of fig.5 \footnote{This
restriction is due to the computation of the Clebsch-Gordan coefficients.}. On
the other hand the $l=\sqrt{S}$-moment decay as a power law $\sim S^{-3/2}$, at
least in the studied range of variation of $S$.

\noindent In view of this, we conclude that, in the GOE case, the critical
value $l_c$ below which the fluctuations are relevant either tends to a
constant, or to infinity slower than $S$, i.e.
\begin{eqnarray}
l_c \sim S^\alpha \ \ \ \ 0 < \alpha < 1
\end{eqnarray}
On the other hand, previous study of the kicked top emphasized the quasifractal
structure of the Husimi density of its eigenfunctions in the chaotic regime
\cite{nak}. This means that fluctuations in both directions of phase space are
present up to the smallest scale allowed by the Heisenberg uncertainty, i.e. up
to
a size $O(\hbar^{1/2})$, which is consistant with eq. (29) with $\alpha=0.5$.
The fact that the moment $\Omega_{\sqrt{S},0}$ also shown on fig.5 decays more
or less as a power-law
\begin{equation}
\Omega_{\sqrt{S},0} \sim 1/S^{-3/2}
\end{equation}
corroborates this conclusion : moments up to $l \sim \sqrt{S}$ are relevant,
i.e. $\alpha=1/2$.

\noindent Nevertheless, nothing forces the  eigenstates of a quantum chaotical
model to match those of a GOE matrix up to the smallest scales. It would
therefore be highly desirable to get a condition on $\alpha$ like eq.(29) for a
quantum chaotical system. This could be achieved by direct computation of
$\Omega_{\sqrt{S},m(S)} $ using eq.(19). However, the numerical difficulty
associated with the computation of high-order Clebsch-Gordan coefficients
renders this task hardly fulfillable, as can be seen on fig.6 where we show
results obtained for $\Omega_{\sqrt{S},0}$ through eq.(19) averaged over more
than 40000 states for each point. On one hand, the semiclassical randomness of
the eigenstates is not attained for small $S$, while on the other hand, the
Clebsch-Gordan coefficients limit the maximal spin magnitude. In other words
these two effects dramatically affects fig.6 left and right. Considering the
size of our statistics, we attribute to these effects the somehow erratic
behaviour of $\Omega_{\sqrt{S},0}$. On fig.6, the solid line indicating a
$S-1.5$-decay is shown as eye-guide, and constitutes in no way a serious
result.

\noindent In conclusion our study of the Husimi density of eigenstates of the
quantum spin system defined by (10) and (11) has confirmed the gaussian shape
of fluctuations around the semiclassical limit. Moreover, the decay of these
fluctuations follow the same power law as in previous studies \cite{eck,fein},
indicating perhaps universality. While GOE results tend to confirm the
quasifractality proposed in \cite{nak}, numerical difficulties forbade us to
check it for the quantum dynamical system. Investigations to overcome this
difficulty are on their way. For the time being, let us just point out that the
fact that GOE eigenstates seem to exhibit this quasifractality renders it a
direct consequence of the randomness of the states. The maximal randomness is
then bounded by Heisenberg's uncertainty, but beside that, the quasifractality
of the states seem to contain no physical content. \\
\\
\\
\noindent We thank the Centro Svizzero di Calcolo Scientifico. Work supported
in part by the Fonds National Suisse de la Recherche Scientifique.

\newpage
\section*{Figure Captions}
\Fig {\bf 1}: Distribution of level curvatures for the eigenstates of (11) and
$S=200 $,  $T=50$ and $\kappa = 1.2$. From the remarkable agreement with RMT
predictions we conclude that the ratio of scarred eigenfunctions is very small
(see \cite{del,tah}), and should therefore not influence our study.\\

\noindent
\Fig {\bf 2}: Moment distribution $P({\hat \Omega}_{1,0}^S)$ as defined in (16)
for a
spin $S=600$. The statistics has been computed from 2404 even states of  four
realisations of (11) taken around $T=50.$ and $\kappa =1.2$. The agreement with
a gaussian (solid line) is remarkable. On inset we show the same curve on a
semi-log plot.\\

\noindent
\Fig {\bf 3}: Log-log plot of the width of the gaussiann distribution of
moments $P({\hat \Omega}_{l,m}^S)$ for model (11), $m=0$ and $l=1$ (squares),
$l=3$
(diamonds) and $l=5$ (triangles) vs. the magnitude of spin $S$. Inset shows the
width of $P({\rm Re}(\Omega_{l,m}^S))$ for $m=2$ and $l=2$ (circles), $l=3$
(squares), $l=4$ (diamonds) and $l=5$ (triangles). In both cases, the solid
line indicates the $S^{-1/2}$ decay.\\

\noindent
\Fig {\bf 4}: Log-log plot of the width of the gaussian distribution of
moments $P({\hat \Omega}_{l,m}^S)$ for GOE, $m=0$ and $l=1$ (circles), $l=3$
(squares)
and $l=5$ (diamonds) vs. the magnitude of spin $S$. The solid line indicates
the
$S^{-1/2}$ decay. \\

\noindent
\Fig {\bf 5}: Log-log plot of the width of the gaussian distribution of
moments $P({\hat \Omega}_{l,m}^S)$ for GOE, $m=0$ and $l=S/2$ (circles), $l=3
S/4$ (squares), $l=S$ (diamonds), $l=5 S/4$ (triangles) and $l=\sqrt{S}$ (empty
diamonds) vs. the magnitude of spin $S$. The upper and lower solid lines
indicate a decay of
$S^{-1.5}$ and $S^{-36}$ respectively. \\

\noindent
\Fig {\bf 6}: Log-log plot of the width of the gaussian distribution of
moments $P({\hat \Omega}_{l(S),0}^S)$ for model (11), $l(S) = \sqrt{S}$ and
$T=50.$ and $\kappa =1.2$, $m=0$ and $S=\sqrt{S}$ vs. the magnitude of spin
$S$. The solid line indicate a decay of $S^{-1/2}$. We attribute the rather
erratic behaviour of the datas to the numerically instable computation of
high-order Clebsch-Gordan coefficients (see text).\\


\newpage
\begin{thebibliography}{99}
\bibitem{eck} B. Eckhardt, S. Fishman, J. Keating, O. Agam, J. Main and K.
M\"uller, Phys. Rev. E {\bf 52}, 5893, (1995).
\bibitem{fein} M. Feingold and A. Peres, Phys. Rev. A {\bf 34}, 591, (1986).
\bibitem{shni} See the addendum of A.I. Shnirelman in :
KAM Theory and Semiclassical Approximations to Eigenfunctions, V.F Lazutkin,
Springer (1993).
\bibitem{hel} E. J. Heller, Phys. Rev. Lett. {\bf 53}, 1515, (1984).
\bibitem{perelo} A. Perelomov, {\it "Generalized Coherent States and their
Applications"}, Springer-Berlin, 1986.
\bibitem{jac} Ph. Jacquod and J.-P. Amiet, J. Phys. A : Math. Gen., {\bf 28},
4799, (1995) and references therein.
\bibitem{del} J. Zakrzewski and D. Delande, Phys. Rev. E, {\bf 47}, 1650,
(1993), and references therein.
\bibitem{tah} T. Takami and H. Hasegawa, Phys. Rev. Lett. {\bf 68}, 419,
(1992).
\bibitem{lili} See e.g. A. J. Lichtenberg and M. A.
Lieberman, {\it "Regular and Chaotic Dynamics"} (Second Edition), Springer,
1992.
\bibitem{amci} J.-P. Amiet and M. Cibils, J. Phys. A : Math. Gen., {\bf 24},
1515, (1991).
\bibitem{boh}P. Pechukas, Phys. Rev. Lett. {\bf51}, 943, (1983) \\ \\
\hspace{1 cm} O. Bohigas, M.-J. Giannoni and C. Shmit, Phys. Rev. Lett.
{\bf52}, 1, (1984)
\bibitem{she} D.L. Shepelyansky, Physica {\bf28}D, 103, (1987).
\bibitem{nak} K. Nakamura, Y. Okazaki and A. R. Bishop, Phys. Rev. Lett. {\bf
57}, 5, (1986).
\end{thebibliography}
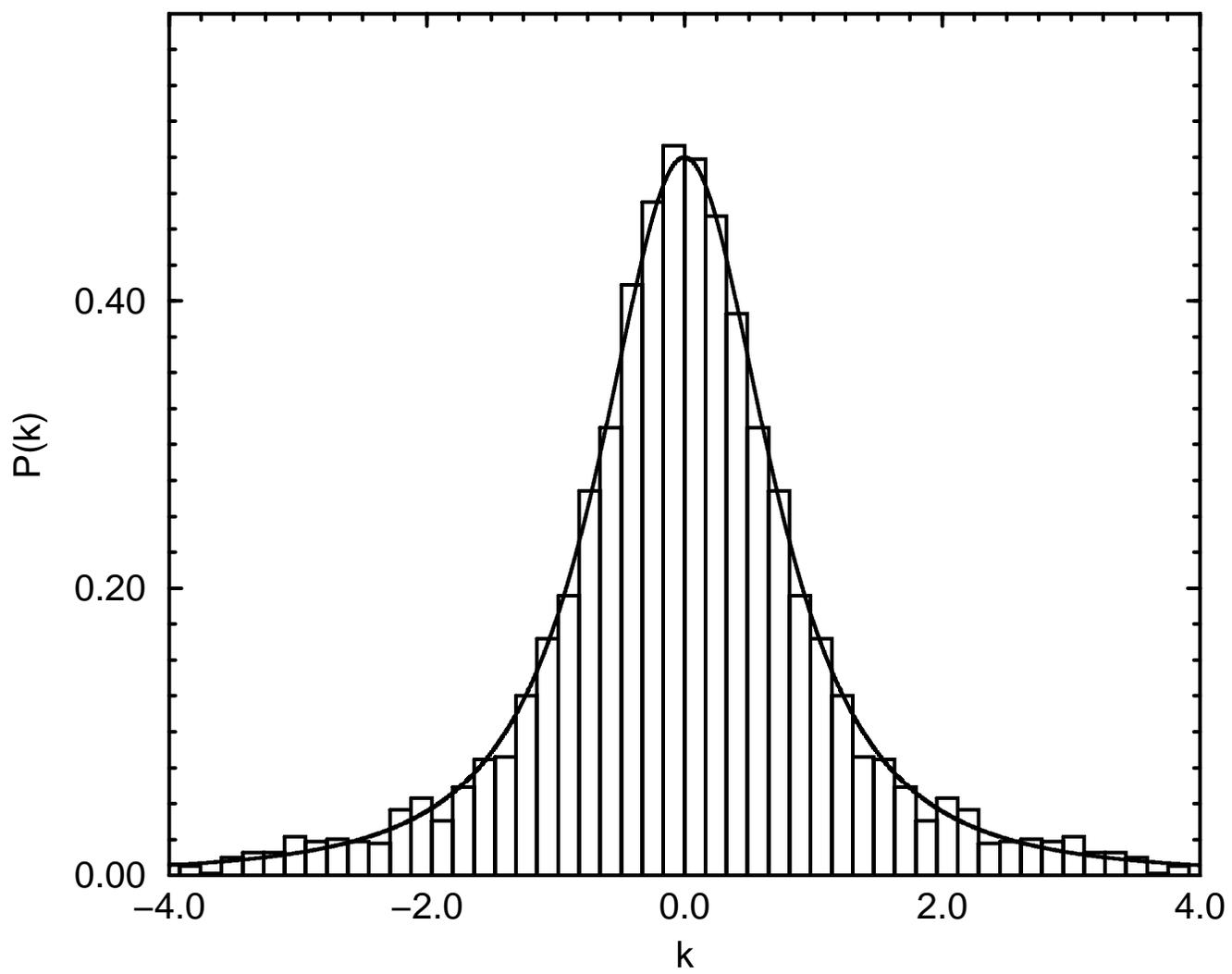
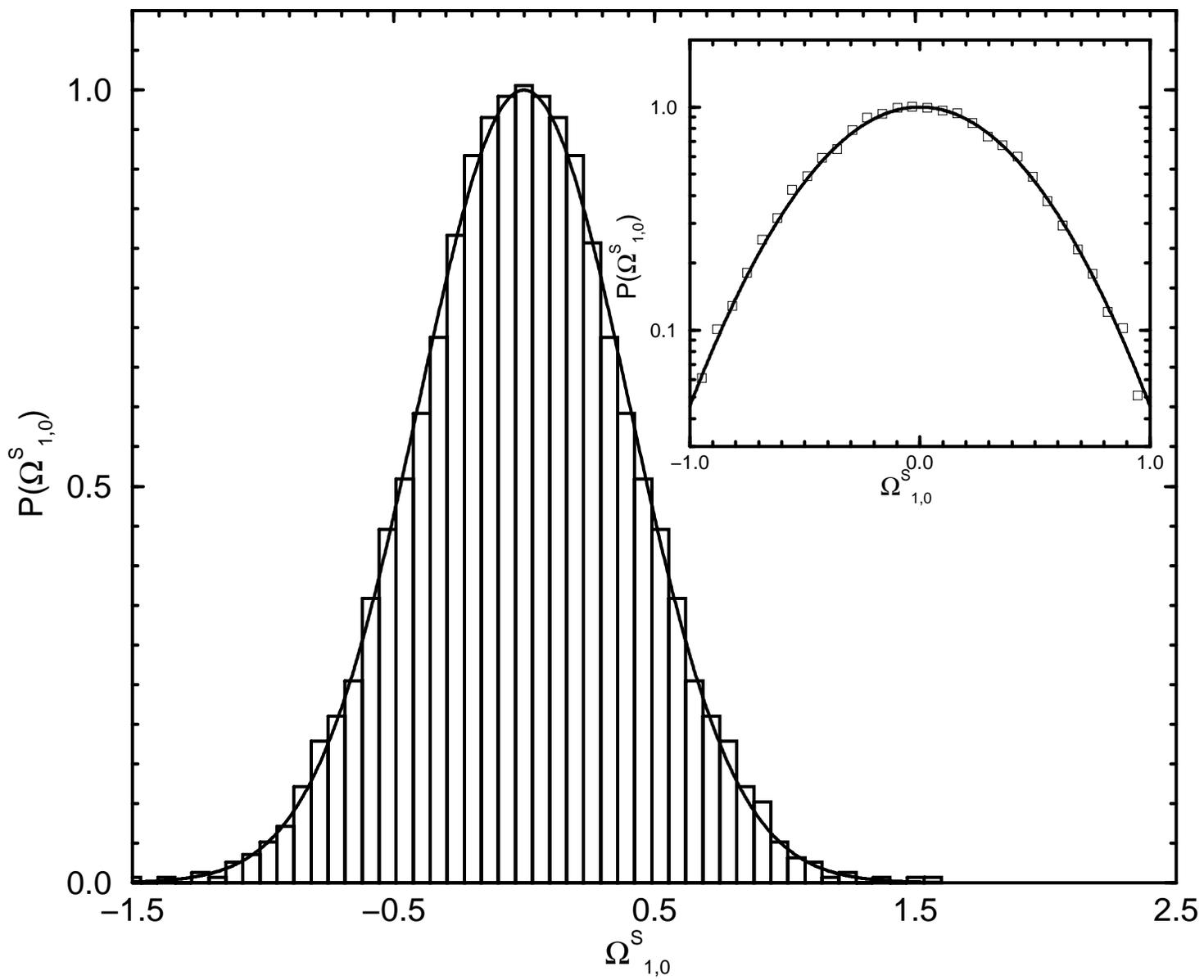
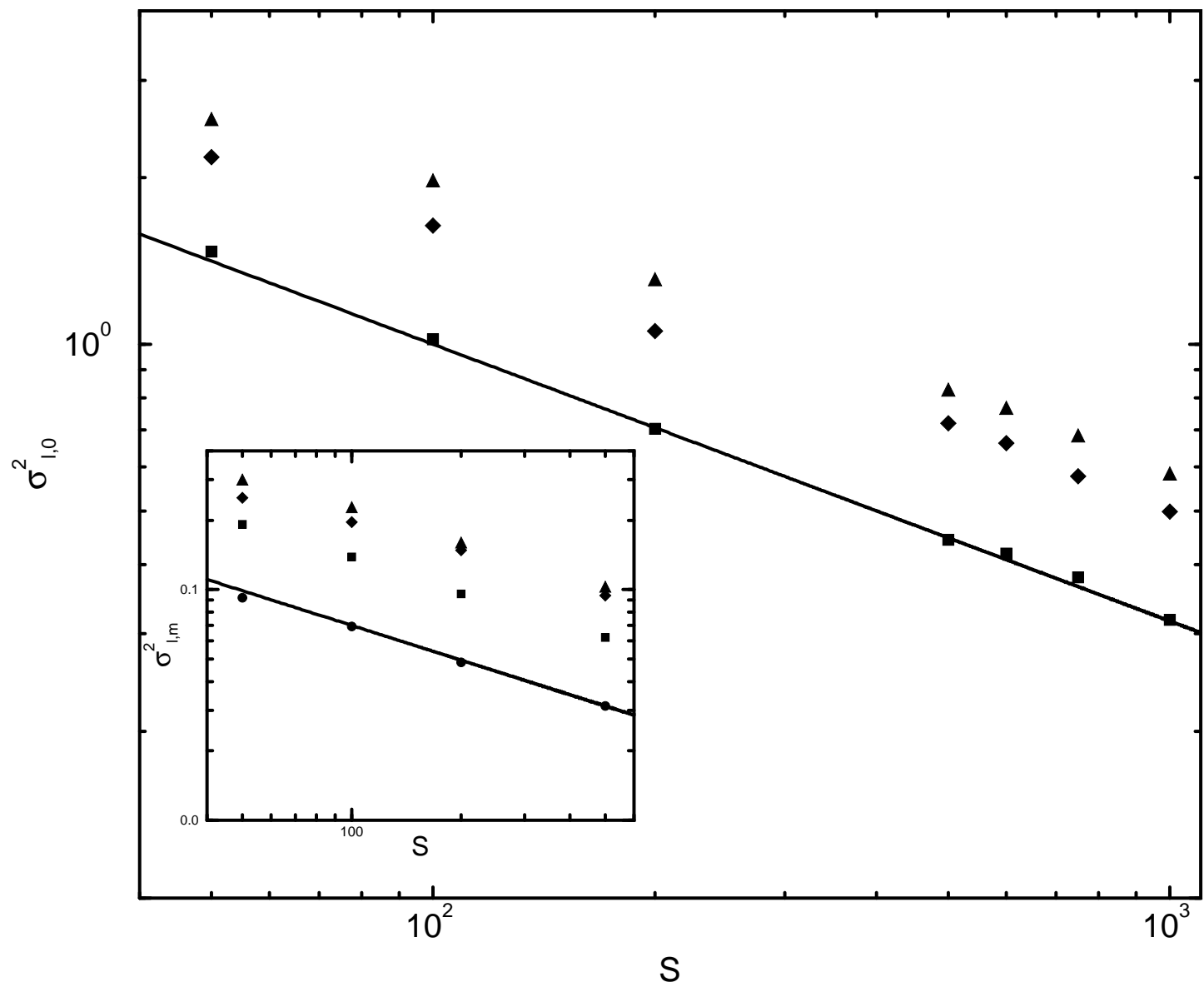
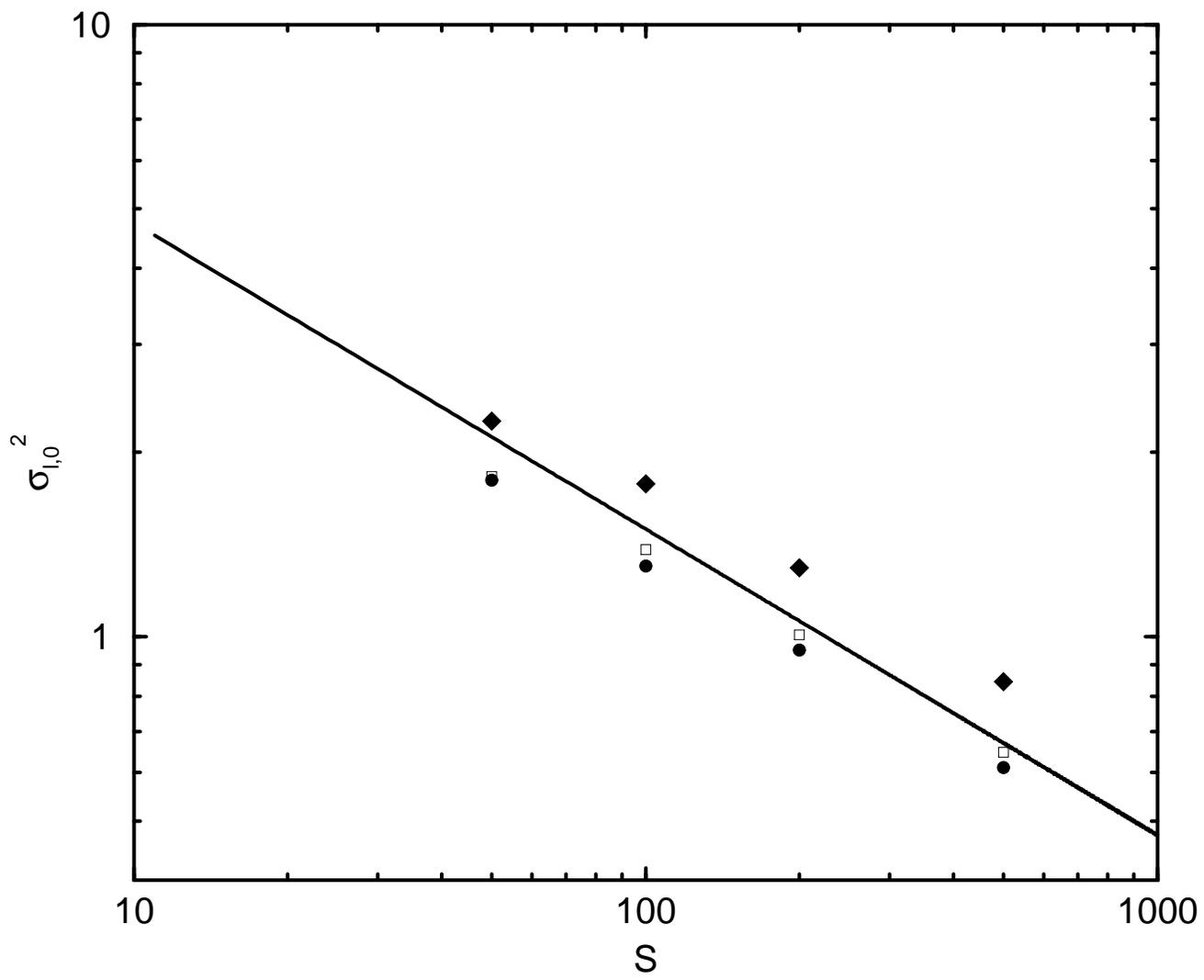
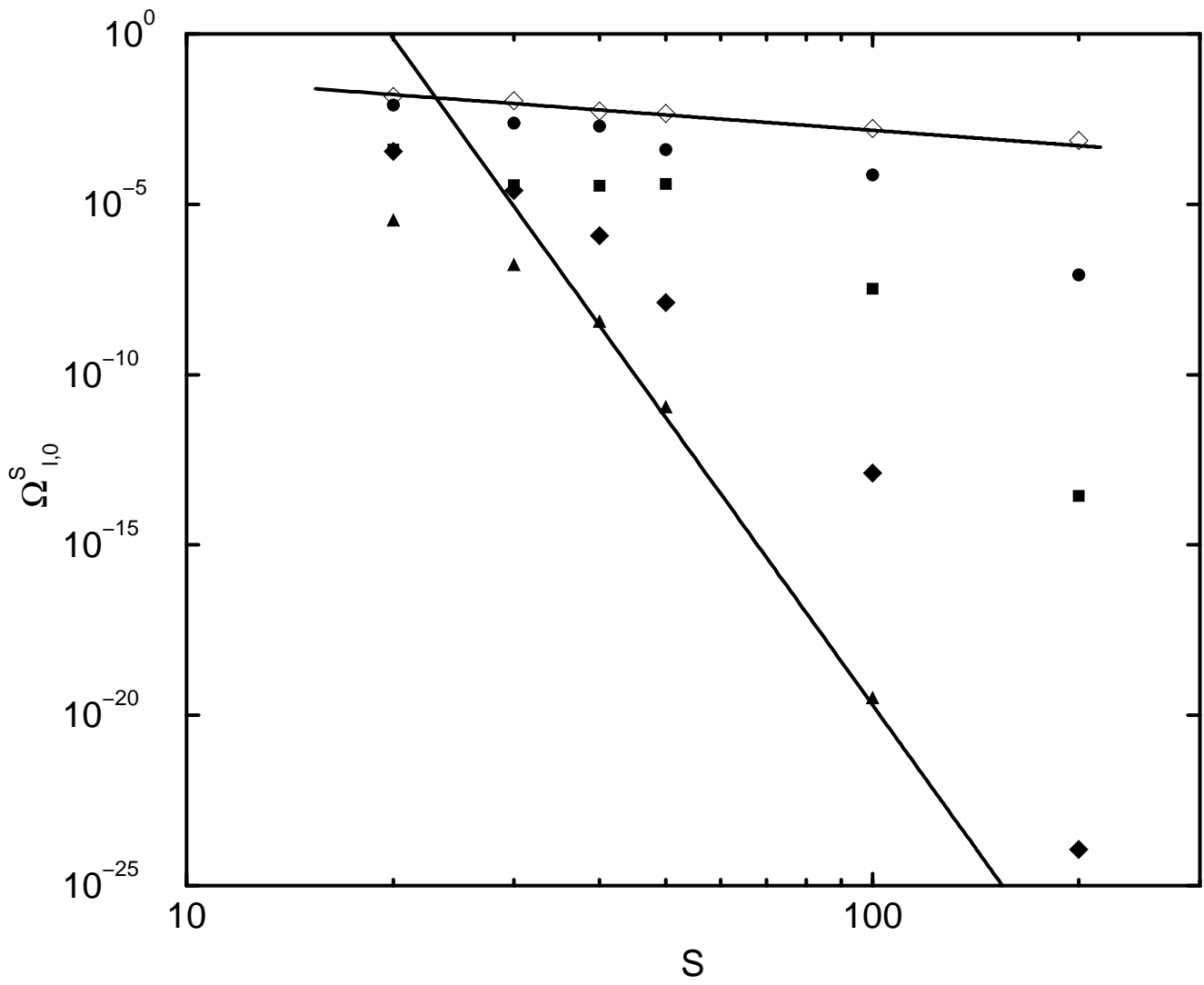
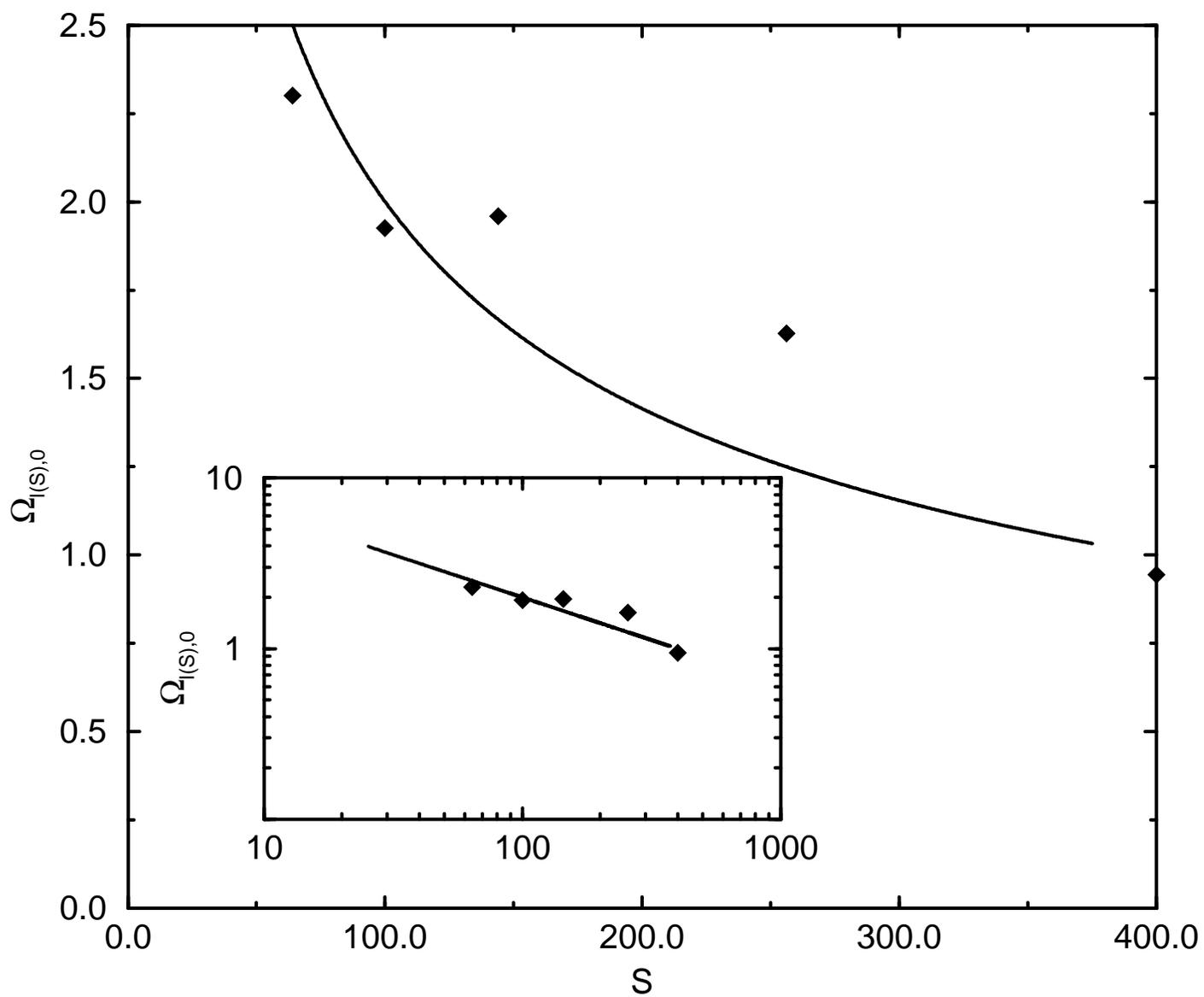
\end{document}